\documentclass{ws-rv10x7}
\usepackage{ws-rv-thm} % comment this line when `amsthm / theorem / ntheorem` package is used
\usepackage[square]{ws-rv-van}

\usepackage{mathtools,dsfont,xfrac, bm}
\usepackage[pdfpagelabels=false,colorlinks=true,allcolors=black]{hyperref} % \label, \ref and \cite are recommended

\renewcommand\thesection{\thechapter.\arabic{section}}
\renewcommand\theequation{\thechapter.\arabic{equation}}
\renewcommand\thefigure{\thechapter.\arabic{figure}}
\usepackage{perpage}\MakePerPage{footnote}

\newcommand \be  {\begin{equation}}
\newcommand \bea {\begin{eqnarray} \nonumber }
\newcommand \ee  {\end{equation}}
\newcommand \eea {\end{eqnarray}}
\renewcommand{\leq}{\leqslant}

\usepackage{amsmath}\DeclareMathOperator{\trace}{tr}

\title{Replica Symmetry Breaking \& Far Beyond} 

\makeindex
\allowdisplaybreaks
\begin{document}
\renewcommand{\thechapter}{1}

\chapter[Simulated annealing, optimization, searching for ground states]{Simulated annealing, optimization, searching for ground states}\label{ch2}
\author[S.~Caracciolo, 
A.~K.~Hartmann, S.~Kirkpatrick, M.~Weigel]{
{Sergio Caracciolo}\footnote{sergio.caracciolo@mi.infn.it}, 
{Alexander K. Hartmann}\footnote{a.hartmann@uni-oldenburg.de},
{Scott Kirkpatrick}\footnote{kirk@cs.huji.ac.il},
and {Martin Weigel}\footnote{martin.weigel@physik.tu-chemnitz.de}}
\address{
\textsuperscript{$\dagger$}Department of Physics, University of Milan and INFN Milan Section, Milan, Italy\\
\textsuperscript{$\ddagger$}
Institut f\"{u}r Physik, Universit\"{a}t Oldenburg, Oldenburg, Germany\\
\textsuperscript{\rm\S}School of Engineering and Computer Science, Hebrew University, Jerusalem, Israel\\
\textsuperscript{\rm\textparagraph}Institut f\"{u}r Physik, Technische Universit\"{a}t Chemnitz, Chemnitz, Germany
}

\begin{abstract}
The chapter starts with a historical summary of first attempts to optimize the spin glass Hamiltonian, comparing it to recent results on searching largest cliques in random graphs.
Exact algorithms to find ground states in generic spin glass models are then explored in Section 1.2, while Section 1.3 is dedicated to the bidimensional case where polynomial algorithms exist and allow for the study of much larger systems. Finally Section 1.4 presents a summary of results for the assignment problem where the finite size corrections for the ground state can be studied in great detail.
\end{abstract}

\newpage
\section{Introduction}\label{sec2.0}

A theme which unites this chapter is the study by simulation of finite random systems in order to test the predictions and insights that arose as new kinds of order were defined and realized in spin glasses. The tools of this chapter have found use in many applied contexts.
In physics typically phase spaces of physical systems have been
considered. In applied mathematics, problems, either idealized or based on
real-world data, are solved to optimize target functions or to satisfy
constraints. Over the years is has been gradually accepted in
mathematics and physics communities that these two realms are very
similar to each other.

Replicas, when introduced by Brout \cite{brout1959statistical}, were large patches of a perfectly ordered lattice. Edwards and Anderson (EA) \cite{edwards:75a} looked for order in the similarity of behavior across multiple identical copies of random spins and their interactions. In the Brout and EA work, the free energy was extracted as the linear term in an expansion of the partition function in powers of $n$, the number of replicas. EA defined order as a self-correlation of spin orientation in different replicas, since this could represent correlation over long times.  Sherrington and Kirkpatrick(SK) \cite{sherrington1975solvable}, after simplifying the problem to an infinite ranged Ising system with i.i.d.\ random Gaussian interactions of strength $1/\sqrt{N}$ in hopes of making it \textit{soluble}, explicitly used the identity,

%Questa equazione non va bene: o si definisce la free energy o si definisce il replica trick
%old Scott
%$F = ln(Z) = lim_{n \rightarrow 0} (Z^n - 1)/n$, 
\begin{equation}
     \ln Z = \lim_{n \to 0} \frac{Z^n - 1}{n}\;,
    \label{da_inserire_dopo_call_con_scott}
\end{equation}
despite the obvious fact that the meaning of $Z^n$ anywhere except at integer values of $n$ was unclear.  
A careful treatment of the extrapolation of the resulting equations for the free energy, $F$, showed an entropy at zero temperature which was negative, an impossibility in a discrete model, while other features such as a  susceptibility cusp seemed plausible.  The predicted ground state energy, $E_0 = (2/\pi)^{1/2} = -0.798...$, could be tested in simulation.

Fortunately, computers in 1975 had almost reached the point where they might be able to address questions about asymptotic values of quantities such as $E_0$. Powerful generally available computers such as IBM's 370-168 and later 3033 could hold as much as 10 MB of data in random access memory, and process instructions at a few MIPS.  (The first supercomputer, the Cray-1, in 1975 also had no more than about 10 MB of RAM, but ran its instructions 4-6 times faster.)  This amount of RAM permitted simulating a small number of samples with 500-1000 spins, so with a floating point coprocessor attached to their IBM mainframe to provide more Cray-like processing speeds, Kirkpatrick and Sherrington (KS) \cite{kirkpatrick1978infinite} in 1978 could report that $E_0$ was in fact between $-0.75$ and $-0.77$, excluding the replica symmetric result. Improvements on that estimate have continued for the next 30+ years. The most recent estimate of the asymptotic result, using Parisi's full RSB formulas and Pad\'e approximants to extrapolate both upper and lower bounds to $E_0$, yields the currently accepted value of $-0.76321...$  \cite{crisanti2002analysis}. The improvements in theory that have made this the accepted result required several decades, and are discussed elsewhere.  Improving the experiments to test it also required some years for computing speeds to increase so that better statistics could emerge from simulations.  Graphs with $N$ up to 2000 sites have been studied \cite{kim2007ground}, but  understanding the size dependence of $E_0$ remains a subject of research.  Simulations agree with theory, and have grown more accurate as the bounds on the theoretical result have also grown tighter. New methods of finding the ground state energies in the simulated model were required, intially and as the models got larger and the accuracy required grew narrower.  These methods, such as simulated annealing, have taken on a life of their own, with many applications in complex systems outside of the simple models of statistical mechanics.

Simulated annealing \cite{kirkpatrick1983} is an obvious idea if you are statistical physicists, like KS when first studying their spin glass model, and also V. {\v{C}}ern{\`y} \cite{vcerny1985thermodynamical} and K. Wilson \cite{KGW}. Kirkpatrick and {\v{C}}ern{\`y} studied the Travelling Salesman as an easily described example.  Wilson used annealing to optimize packing of parallel processor code generated in a compiler he wrote for his quantum chromodynamics simulations. All of us realized that allowing a Markov chain of states to evolve at a series of decreasing finite temperatures (\textit{annealing}) would tend to reach larger and deeper minima of a complex function space with many local minima than simple gradient descent. 

At the time the preferred methods used gradient descent with many random restarts and tricks to jump to new starting points, applied only when a search gets stuck. The large number of possible local minima, mostly at higher energies, makes this ineffective.  In addition, study of properties such as susceptibilities and specific heat (obtained from the fluctuations of a Boltzmann system) could guide the development of effective annealing schedules.  Ranges of temperature at which large fluctuations were seen, possibly signalling a phase boundary, give a signal to slow the annealing process to allow large changes to occur.

Kirkpatrick and IBM colleagues \cite{kirkpatrick1983}  were also at the time studying the problems presented by design automation tools used to place and route transistors and small VLSI circuits in IBM's next-generation computers, and included simplified examples of both practical problems in their paper.  These tools were successfully used and incorporated into IBM's internal product development tool sets, and adopted in other industries. 

The IBM group learned important lessons from exposure to
the engineering environment.  First, expressing complex constraints as energetic costs in a Hamiltonian framework provided valuable flexibility, and the annealing framework met the constraints simultaneously or in order of importance rather than one at a time. Second, they soon realized that
finding the exact optimum in a problem like circuit design was not really the objective of an engineering team, who simply needed to find solutions that were good enough, soon enough, to ship as  products.  For that sort of objective, robustness of a methodology and the ability to incorporate many esoteric constraints was at least as important as its efficiency. Optimality of the solution obtained, if possible, was a bonus, but not the only objective.

Simulated annealing has been accepted as another tool for at least approximately solving messy but useful and important problems. Probably its greater impact has been to stimulate the incorporation of statistical physics frameworks as an active part of applied mathematics.  One recent review by a respected practitioner of constraint satisfaction \cite{Selman} considers the 1990s to be the era of phase transitions, with the 2000s introducing more refined techniques such as survey propagation \cite{mezard2002analytic} and cavity constructions \cite{mezard2003cavity}. Phase diagrams with pictures of changes in the shapes of possible sets of solutions and different degrees of ergodicity in the phase space of a problem's solutions \cite{krzakala2007gibbs,zdeborova2009statistical} now shape the discussions of the fundamental differences between different complexity classes of combinatorial and computational problems \cite{gamarnik2021overlap}.

\begin{figure}[tb!]
\centering
\includegraphics[width=1\columnwidth, keepaspectratio=true, angle=0]{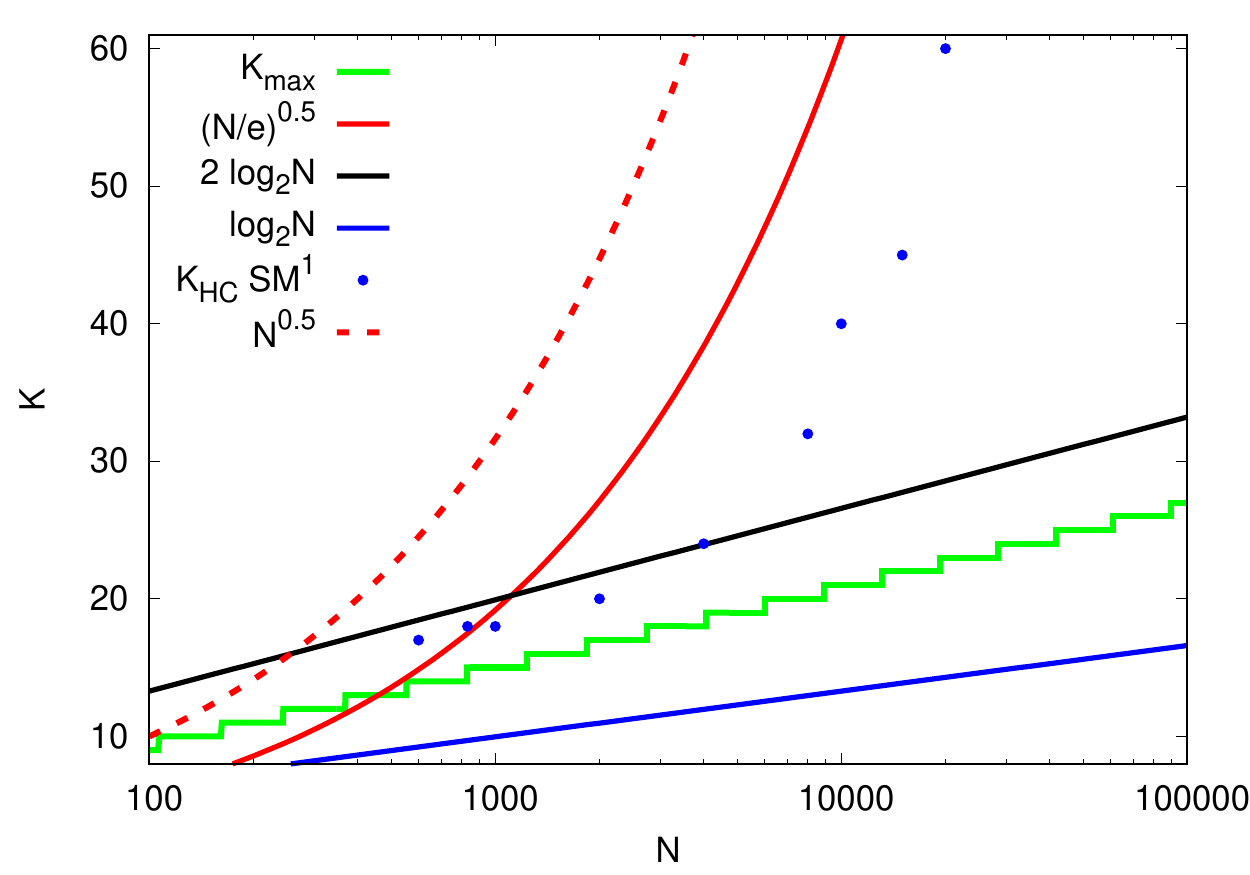}
\caption{ Hidden clique sizes $K_{HC}$ of interest in $G(N,p=0.5)$ lie between the $K_{\text{max}}$ staircase and the proven or experimentally observed lower limits that can be found with spectral methods (dashed red line) or message-passing techniques (solid red line).  Sizes of the smallest hidden cliques identified by a simple greedy search from all initial sites, stopping as soon as evidence for the hidden clique is found, are shown with blue dots and lie well below both limits.}
\label{fig-mont}
\end{figure}

The simplest problems in combinatorial optimization are proving explicitly some property of a random graph, or proving that the property doesn't hold in that graph.  An example would be testing if the maximum size of a clique, or completely connected cluster of sites, of size $K$ exists in an Erd\H{o}s-R\'enyi graph $G(N,p=0.5)$ of size $N$, with half of the bonds, selected at random, the rest absent.  This is a stylization of a common question data scientists might ask of the social networks in tracking interaction on today's network, where such graphs may describe populations of billions, and the bonds might be demonstrated common interests or communication. We know what is possible by just calculating expectations. The largest cliques, of size $K_{\text{max}}$, that form naturally in this model will not exceed $2\log_2 N$ in size, with finite-size corrections making the actual limiting size of such a \textit{MaxClique} much less.  Simple linear algorithms, with costs proportional to the number of bonds, thus $N^2$, fail to find solutions with clique size $K$ more than a few sites bigger than $\log_2 N$.  A staircase of steps at which increasingly larger maximum cliques are found thus bounds a \textit{hard} phase in which large numbers of cliques with sizes greater than $\log_2 N$ must exist, but cannot be found without introducing some more expensive nonlinear search, such as backtracking.

A popular extension of this problem is to add a \textit{hidden clique} to the random graph which is larger than those which occur at random. Since this clique is unique, it should be easier to find, but the best methods proven to work with quasi-linear cost (but large prefactors) can only find hidden cliques of size greater than $\sqrt{N/e}$.  So an easy region for this related problem exists when the hidden clique is bigger than this line.  The resulting phase diagram is shown in Fig. \ref{fig-mont} \cite{marino2022hard}.

Some recent work has returned to the original intuition of EA that correlations of activity between identical replicas of random, symmetry-less systems were an  indication of ordering that might serve in place of a traditional order parameter.  Consider simulating the evolution of multiple copies of a random system at steadily lowered temperatures, or at a number of different temperatures, and passing information about improved solutions between the different replicas, a methodology called \textit{parallel tempering} in physics \cite{hukushima1996exchange}.  This has proved capable of finding hidden cliques in very large graphs when the hidden clique barely exceeds the size of naturally occurring cliques \cite{chiara2018parallel}, as well as to find maximum independent sets very efficiently \cite{angelini2019monte}.  Annealing multiple replicas in this way has the disadvantage that it cannot be proven to have only polynomial cost (in the size of the problem) asymptotically, although experience and empirical knowledge of the problem may allow tuning the method for affordable costs on large practical problems. It may also provide a path to overcoming the confusion that impedes the search for optimal configurations in combinatorial problems where many suboptimal configurations overlap and confuse any local search, such as coloring or maximum clique \cite{marino2022hard}. 

The extra search power that parallel tempering seems to provide suggests that there may be other ways to exploit the original insight of Edwards and Anderson, that ordering in random systems might show up as a correlation between behavior observed in multiple copies or regions of a complex system.  Replica symmetry breaking implies ordering which may take on a range of values, and study of the correlations seen in multiple replicas at low temperatures may add physical insight into such ordering.  A further step could be to model systems which are basically similar but subject to weak local environmental corrections with a set of replicas, each slightly different.  The objective would be to identify from the differences between solutions in each replica a functional variation of the response of ordering to the different local influences.  Such local variations are typical of much of the vast amounts of social data available for study in today's highly instrumented world.
This is just one more possible, but as yet unrealized, impact of spin glasses on applied mathematics and statistics.  

The original SK Ising model with Gaussian-distributed interactions is only one of the situations which has stimulated further exploration to tease out the lowest energy ground states. Different ways of combining different tools of optimization have proved valuable in this and different lessons emerge, as covered in the further sections of this chapter.

\section{Focusing on Ground States}\label{sec2.1}

Consider the spin-glass (SG) model in zero field with Hamiltonian
\cite{edwards:75a}
\begin{equation}
  {\cal H} = -\sum_{\langle i,j\rangle} J_{ij} s_i s_j.
  \label{eq:hamiltonian}
\end{equation}
We focus on Ising spins $s_i = \pm 1$ placed on the vertices of a
graph, but other symmetries of the order parameter
have been considered. The Ising symmetry has the advantage that the
related statistical physics problems are combinatorial.
The sum $\langle i,j\rangle$ runs over all edges of the graph.
For the Edwards-Anderson (EA) model the graph is a hyper-cubic
lattice in $d$ dimensions, but also other graph structures
like the fully connected Sherrington-Kirkpatrick model corresponding
to the $d\to\infty$ limit of \eqref{eq:hamiltonian} have been studied.
As discussed in Chapter 1,
the physics of this mean-field problem is
well understood, in many aspects since the 1980s. The situation in low dimensions is
less clear and so much of the work in recent decades has focused on studying the
problem in 2, 3 and 4 dimensions with numerical methods.

Next to Monte Carlo simulations used to study the vicinity of the spin-glass
transition and the behavior in the ordered phase (see Chapter 5), much work has been
invested in studying ground states and low-lying excitations above them in order to
gauge the low-temperature behavior. Alternative theories for the nature of the
spin-glass phase make contrasting predictions about the energetic and geometric
properties of such excitations. Based on generalizations of Peierls' argument for the
stability of the ferromagnetic phase, researchers have argued that excitations with
energies diverging with length will lead to spin-glass phases that are stable against
thermal fluctuations, such that one of the prime goals of studying ground states for
spin-glass systems is to understand the nature of the prevalent low-energy
excitations and how their sizes and energies are related.

The search for ground states for spin glasses with discrete symmetry amounts to a
combinatorial optimization problem as there is a countable number of candidate
solutions \cite{hartmann:book}. Such problems can hence be solved by a brute-force
enumeration of all configurations, with an effort that grows exponentially with the
number of spins. As discussed in section \ref{sec:SG:alg:exact}, 
a clever organization of this
enumeration can lead to massive gains in efficiency of this process, but will not
alter the overall exponential scaling of the process. 
This corresponds to the fact that the
problem in $d \ge 3$ is known to be {\em NP\/} hard \cite{barahona:82}.
In some special cases, however,
the ground-state problem can be mapped onto auxiliary optimization problems that
permit solutions in {\em polynomial\/} time. This will be explained in section
\ref{sec:SG:alg:poly}. Beyond the ground states of spin glasses, many optimization problems
have caught the attention of physicists \cite{Hartmann2006book,Mezard2009information}. 
An example for a recently studied model
is the assignment problem, which is discussed in section \ref{sec:assignment}.

\section{Exact Algorithms for hard problems \label{sec:SG:alg:exact}}

Here, exact and general algorithms for 
finding ground states (GSs) for Ising SGs are considered,
although the presented methods are very general.
As an extension of Eq.~(\ref{eq:hamiltonian}), here
also an interaction with local fields $h_i$ is included. 
For most studied systems no local field is present, but for some
of the algorithms presented below, such a term might arise
for sub-problems to be solved. 
In this case, the energy of a SG is given by the Hamiltonian
\begin{equation}
{\cal H}(s)= - \sum_{i<j} J_{ij} s_i s_j\, - \sum _i h_i s_i \label{eq:SG0}\,,
\end{equation}
where $s=(s_1,\ldots,s_N)$ is a configuration and the bonds $J_{ij}$,
corresponding to the edges of the graph, 
are quenched random variables. The sum runs here,
in the most general case, over
all pairs of spins. Thus, if all bonds are nonzero, the system
is of mean-field type. For finite-dimensional lattices, most bonds
are zero, suitably selected. The nonzero bonds
are typically drawn from a Gaussian or a bimodal $J_{ij}=\pm 1$ distribution.

 Here, three types of algorithms are
presented, which have frequently been used to calculate
exact GSs for three-dimensional SGs. First,
the \emph{branch-and-bound} method is explained, which is based
on enumerating many states in a sophisticated way. Next, the
\emph{linear programming} approach is outlined, which is based
on rewriting the Hamiltonian as a linear function, relaxing the
condition $s_i=\pm 1$, plus adding additional constraints, called
\emph{cutting planes} or \emph {cuts}. 
Finally, the combination of both approaches,
the \emph{branch-and-cut} algorithm is presented, which yields the
currently fastest exact method to obtain spin-glass ground states.

\subsection{Branch-and-bound}

The basic idea of the branch-and-bound approach \cite{land1960} 
to find the minima
of  Eq.~(\ref{eq:SG0}) is to represent all spin configurations
as a binary tree, where at each node the configuration space
is, for a node-dependent selected spin $i_0$,  
subdivided into the configurations with $s_{i_0}=+1$ and
those with $s_{i_0}=-1$, respectively. The spin configurations
are given by the $2^N$ leafs of this tree.
The simplest approach to the GS problem would be to obtain all 
configurations by enumeration, and simply pick those with the
minimum energy, yielding an $O(2^N)$ running time.

This running time can be improved, although still being exponential
in the worst case,
by omitting parts of this tree via considering \emph{bounds} on
the achievable energies in sub-trees \cite{kobe1978}. Here
we present the refined algorithm described in Ref.~\cite{hartwig1984}.
The branching is performed always on the last spin of the sub-problem,
i.e., it starts with spin $N$. The energies for the sub problems
with $s_N=+1$ and $s_N=-1$ can be written using Eq. (\ref{eq:SG0})
for $\overline{s}=(s_1,\ldots,s_{N-1})$ as follows:
\begin{eqnarray}
{\cal H}^+(\overline{s}) & = &
-{\sum_{i<j}}' J_{ij}s_is_j - {\sum_i}' h_is_i -{\sum_i}' J_{iN}s_i-h_N \;,
\label{eq:Hplus}
\\
{\cal H}^-(\overline{s}) & = &
-{\sum_{i<j}}' J_{ij}s_is_j - {\sum_i}' h_is_i +{\sum_i}' J_{iN}s_i+h_N \;,
\label{eq:Hminus}
\end{eqnarray}
where the sums $\sum'$ run from $1$ to $N-1$. If one defines 
\begin{equation}
{\cal H}^*_{N-1} = \min_{s_1,\ldots,s_{N-1}} -{\sum_{i<j}}' J_{ij}s_is_j\;,
\end{equation}
one obtains, using the relation 
$\min_s(f_1(s)+f_2(s))\ge \min_s f_1(s) + \min_s f_2(s)$,  the bounds
\begin{eqnarray}
{\cal H}^+(\overline{s}) & \ge & {\cal H}^* + \min_{s_1,\ldots,s_{N-1}}
 {\sum_i}' (-h_i-J_{in}) s_i -h_n \;,\\
{\cal H}^-(\overline{s}) & \ge & {\cal H}^* + \min_{s_1,\ldots,s_{N-1}}
 - {\sum_i}' (-h_i+J_{in}) s_i +h_n \;.
\end{eqnarray}
These bounds are available because
the minima ${\cal H}^*_{N-1}$, ${\cal H}^*_{N-2}, \ldots$  can be obtained recursively
while the branching tree is built. Furthermore, the minimum of a linear
function $\sum_s \sum_i a_i s_i$, here $a_i=-h_i-J_{in}$ or
$a_i = -h_i+J_{in}$, respectively, is simply given by $-\sum_i |a_i|$.
Thus, as a first way to restrict the size of the branching tree,
if one or two of the branches exhibit bounds above a known threshold,
the corresponding branches can be omitted. Such thresholds may come from
low lying configurations found already in other branches or by heuristic
algorithms, or simply given as part of the problem 
in case an enumeration of a specified range
of low-lying  configurations is sought.

A second type of bound \cite{hartwig1984} works as follows. Using 
$d(\overline s) = -2{\sum_i}' J_{iN}s_i-2h_N$ one can rewrite
Eqs.~(\ref{eq:Hplus}) and (\ref{eq:Hminus}) as  
${\cal H}^+(\overline{s})={\cal H}^-(\overline{s})+2d(\overline{s})$. Therefore,
we obtain the implications
\begin{eqnarray*}
\max_{s_1,\ldots,s_{N-1}} d(\overline{s})= 2 \sum_i |J_{iN} |-2h_N \le 0
& \,\Longrightarrow\, & {\cal H}^+(\overline{s}) \ge {\cal H}^-(\overline{s})\;, \\
\min_{s_1,\ldots,s_{N-1}} d(\overline{s})= -2 \sum_i |J_{iN} | -2h_N \ge 0
& \,\Longrightarrow\, & {\cal H}^+(\overline{s}) \le {\cal H}^-(\overline{s}) \;.
\end{eqnarray*}
These bounds are easy to calculate and allow sometimes for omitting
one of the two branches, before any branch has to be evaluated.

Such branch-and-bound algorithms have been used, e.g., to analyze
the low-temperature landscape 
\cite{klotz1998,klotz1998b,krawczyk2002,schubert2006}  of SGs.
% NOT \cite{seyed-allaei2008} because they enumerate all states
More broadly, in the field of statistical mechanics of optimization problems
\cite{Hartmann2006book,Mezard2009information},
the branch-and-bound approach has been applied to other
problems like the satisfiability
problem \cite{monasson1999} or the vertex-cover problem \cite{cover2000}.
For the latter one, a branch-and-bound approach was used, where the variable
to branch on was not selected in a given order but determined
by a local heuristics for further reduction of the branching tree.
For simple variants of these algorithms, it has also been possible to calculate analytically the typical
running time for ensembles of random problems, which exhibit
transitions between typically polynomial and typically exponential
behavior \cite{cocco2001,cover-time2001}.

\subsection{Linear programming and cutting planes}

A different approach works by translating the quadratic Hamiltonian
into a linear problem.
For convenience, here we consider the form of Eq.~(\ref{eq:hamiltonian})
Note that the field term present in Eq.\ (\ref{eq:SG0}) can be written
as quadratic term by introducing a ``ghost spin'' $s_0=1$ matching the
given equation.

We describe the system by a graph $G=(V,E)$ where $V$ denotes the set
of sites where the spins are located and $E$ the set of edges $\{i,j\}$
between the interacting sites.
For any set $V'\subset V$ in $G$
the \emph{cut} $\delta(V') \subset E$ 
denotes the set of edges where one endpoint
is in $V'$ and the other is not, i.e., those connecting the two
sets $V'$ and $V\setminus V'$.
For each configuration, the spins can be partitioned into two sets 
$V^+=\{i|s_i=+1\}$ of ``up'' spins and $V^+=\{i|s_i=-1\}=V\setminus V^+$ 
of ``down'' spins. If two spins $s_i$ and $s_j$ are oriented identically, 
they contribute the energy
$-J_{ij}=+0-J_{ij}$, while they contribute the energy $J_{ij}=2 J_{ij}-J_{ij}$
if they are oriented differently. Thus, using the cut, we can write
\begin{equation}
{\cal H}(s) = 2\sum_{\{i,j\}\in\delta(V^+)} J_{ij} - \sum_{ij} J_{ij}\,.
\label{eq:SG2}
\end{equation}
By introducing $c_{ij}=-J_{ij}$, $S=\sum_{ij} J_{ij}$ and variables
$x_{ij}=1$ if $\{i,j\}\in\delta(V^+)$ and $x_{ij}=0$ else, this reads
\begin{equation}
{\cal H}(x) = -2\sum_{i,j} c_{ij}x_{ij} - S\,, \label{eq:cut:weight}
\end{equation}
where $\sum_{i,j} c_{ij}x_{ij}$ is called the \emph{weight}
 of the cut represented by 
$x$. Therefore, since $S$ is only a constant, 
finding the \emph{minimum energy} of Eq.~(\ref{eq:hamiltonian}) corresponds to finding
the maximum cut in a graph with edge weights $c_{ij}$.
Note that Eq.~(\ref{eq:cut:weight}) is a \emph{linear function} with
integer variables, which means
we have transformed the quadratic optimization problem into a linear one,
but with additional constraints since $x$ must describe a cut.
This problem is called \emph{integer linear program}.

To include the constraints describing the cut, we note that for any cycle 
in the graph $G$ the cut must be crossed an even number of times.
This  can be conveniently described by linear inequalities
\cite{barahona1986} for the variables $x$ 
as follows: For any cycle $C\subset E$, i.e.,
a path which starts and ends at the same vertex, and any odd cardinality
subset $Q\subset C$,
\begin{equation}
\sum_{\{i,j\}\in Q }x_{ij}-\sum_{\{i,j\}\in C\setminus Q }x_{ij} \le |Q|-1
\label{eq:constraints}
\end{equation}
must hold, which defines a cut comprehensively.

The basic idea of the \emph{cutting-plane} approach is now to \emph{relax}
the variables $x_{ij}=0,1$ to $0\le x_{ij}\le 1$ and look
for a maximum of Eq.~(\ref{eq:cut:weight}) given the linear inequalities 
Eq.~(\ref{eq:constraints}). The name cutting plane
comes from the fact that all inequalities describe hyper-planes which
cut off one part from the space of possible solutions.
The resulting problem is called a \emph{linear program} (LP). 
The good
news is that
LPs, i.e., with the relaxed variables, can be solved \cite{padberg1995} 
in worst-case polynomial time in the problem size 
using the \emph{ellipsoid method}. Still, in a practical context
methods like the \emph{simplex approach} \cite{dantzig1948} 
or the \emph{dual simplex
approach}, which exhibit no polynomial bound, perform much faster.
Still, the bad news is that the number of inequalities is in principle
exponentially large, which would yield an exponential running time
right-on. Therefore, instead of adding all constraints immediately
to the LP, one starts with no or few constraints and calculates a first solution. 
Since fewer constraints than necessary are contained in the relaxed
problem, typically the obtained cut weight will be higher, 
i.e., an upper bound of the true maximum cut for the integer LP.
Hence, 
the obtained solution will typically not correspond to a cut and not
be pure integer-valued.
Thus, there may be some of the exponentially many 
inequalities which are violated. The second basic idea
of the cutting plane approach is to look specifically for violated inequalities,
without searching all possible ones,
and add the violated ones to a growing set of inequalities included in the LP.
An exact polynomial algorithm to find violated inequalities
is based on solving a series of shortest-path problems \cite{barahona1986},
which can be done in $O(N^3)$. This is polynomial but rather slow. 
Fortunately, there
exist, in particular for physical lattice structures, several
heuristics which run in linear $O(N)$ time \cite{liers2004} and are most of the
time sufficient to generate additional inequalities for violated conditions. 
If at some point no further violated inequalities are found and the 
solution is fully integer-valued, a true optimum has been found, and the algorithm 
stops.
But if on the contrary  some variables are
still non-integer,  one can resort to branching as 
explained in the next subsection.

Nevertheless, for some combinatorial problems it has been observed that
a cutting plane approach alone may lead to a valid optimum
solution of the integer problem. For example for 
vertex-cover problems on Erd\H{o}s-R\'enyi random graphs
a different kind of cycle inequalities has been considered  \cite{vc_lp2012}.
When varying the average number $c$ of neighbor nodes in the graphs,
a phase transition has been observed. In the thermodynamic limit $N\to\infty$, 
for small values 
$c<c^*=e\approx 2.71$, typically all instances can be solved completely
in a polynomial running time,
while for larger values of $c$ this is not possible. Interestingly,
this \emph{easy-hard transition} coincides with the critical
connectivity where replica-symmetry breaking of the vertex-cover
problem occurs \cite{cover2000}. At this point
a complex structure of the solution space has been observed
\cite{vccluster2004}, where the so-called \emph{leaf-removal core}
\cite{bauer2001} starts to percolate.

Finally, note that there is a set of cutting planes, so called
\emph{Gomory cuts} \cite{gomory1958}, which can be constructed generally for 
all relaxed linear problems. They are based on identifying
violated inequalities directly from a given non-integer
solution, actually in the so-called \emph{tableau} \cite{padberg1995}
used by the simplex algorithm to solve the LP. 
It is proven that this, possibly exponentially large,
set of inequalities is complete, i.e. \emph{any} integer programming
problem can be solved in principle by generating \emph{just} 
these cutting planes. Still, in
practice, any finite numerical accuracy leads to convergence problems,
such that this and many other cutting-plane approaches turned
out to be actually
efficient in combination with branching, as explained next.

\subsection{Branch-and-cut}

If for a relaxed linear system describing maximum cuts
 the solution is still non-integer,
one considers for some variable $x_{ij}\neq 0,1$ both possibilities
$x_{ij}=0$ and $x_{ij}=1$ and solves the corresponding sub-problems
recursively. This means one branches. The combination of these
approaches is therefore called \emph{branch-and-cut}
\cite{barahona1988,liers2004}. Note that
here several other bounds, in addition to those mentioned above,
can be used. This can be lower bounds obtained from the cut weight
of any valid  cut $x^*$ or upper bounds obtained from
suitable relaxations like the LP.

Currently, the branch-and-cut approach can be considered
as the most powerful exact algorithm to obtain GSs
for hard SG instances. 
A publicly accessible implementation of a branch-and-cut algorithm
is the \emph{spin-glass server} \cite{sg-server}
hosted by the University of Bonn. The server was originally implemented 
at the University of Cologne by several members and
collaborators of the
research groups of F. Liers 
and M. J\"unger. At the server, you can submit SG
instances as a file and, if the system is not too large,
 a corresponding GS will be returned.

Branch-and-cut approaches have been applied to finite-dimensional
SGs in several occasions. To our knowledge the largest
three-dimensional instances were $N=12^3$ as considered in a study of low-lying
excitations \cite{palassini2003}.

The approach has also been applied for a mean-field 
random-bond Ising model, which is a generalization
of the standard SG with a variable fraction of negative bond
as controlled by a non-zero mean of the bond values. Here, 
a phase transition of the typical branch-and-cut running time,
as measured by the number of LPs solved,  
between an easy and a hard phase has been observed near the
transition between ferromagnetic and SG behavior 
\cite{viana_bray2003}.

\section{Ground states in two dimensions\label{sec:SG:alg:poly}}

We now turn to the case of ground-state problems permitting a polynomial-time solution. For the EA model of Eq.~\eqref{eq:hamiltonian} this is the case in dimensions $d < 3$.  Since the 1d problem is rather trivial, the much more interesting case of this type is the EA
model in two dimensions.

\begin{figure}[tb!]
  \begin{center}
    \includegraphics[width=0.95\textwidth]{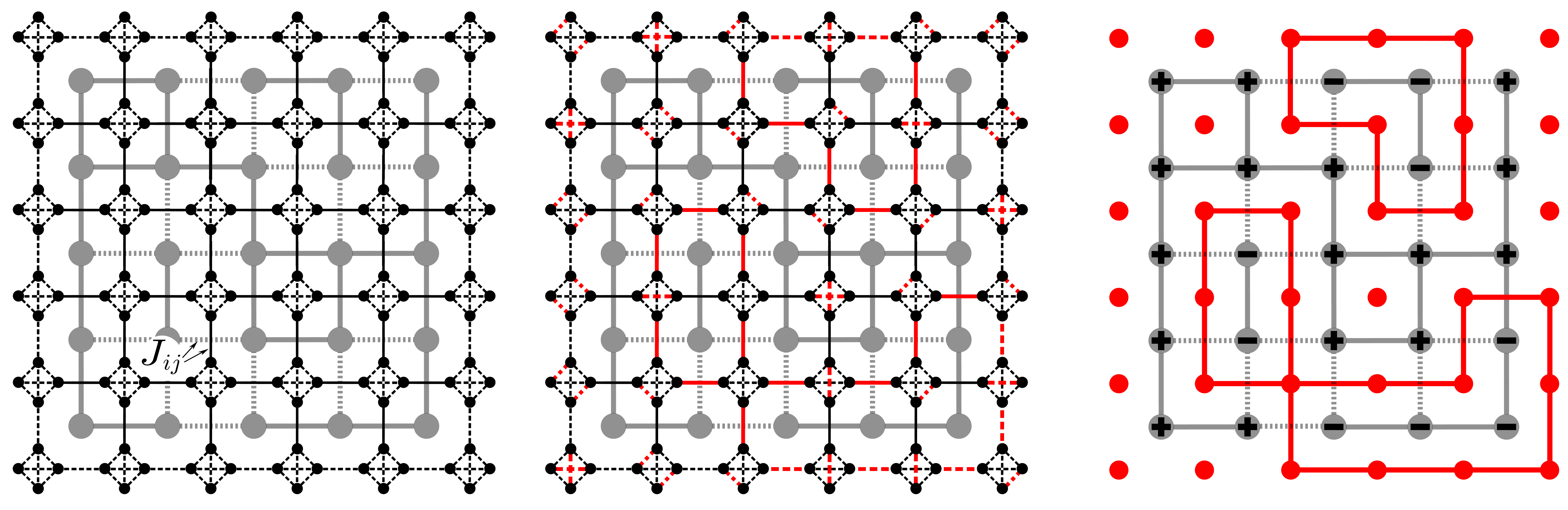}
  \end{center}
  \caption{Mapping of the Ising ground-state problem to a matching problem on an
    auxiliary graph with Kasteleyn cities. Left: expansion of the Ising lattice,
    replacing each vertex by a complete graph $K_4$. Edge weights in the $K_4$
    subgraphs are set to 0, the remaining weights are  $J_{ij}$. Middle: A minimum-weight perfect matching on the auxiliary
    graph. Right: Back-transformation from the decorated graph to the original
    lattice. The matching then results in a set of closed loops separating up from
    down spins.}
  \label{fig:matching}
\end{figure}

A relevant mapping of the EA ground state to a minimum-weight perfect matching (MWPM)
problem on an auxiliary graph was first proposed in Ref.~\cite{bieche:80a}. It is based on
the observation that frustrated plaquettes \cite{toulouse:77a}, i.e., elementary
lattice faces including an odd number of antiferromagntic couplings, must have an odd
number of broken bonds with $J_{ij}s_i s_j < 0$, while non-frustrated plaquettes have
an even number of broken bonds. Hence a spin configuration can be depicted as a
configuration of defect lines of broken bonds that start and end at frustrated
plaquettes. A ground state configuration is then a perfect pairing (matching) of
frustrated plaquettes through defect lines of minimum total weight. MWPM can be
solved in polynomial time based on the blossom algorithm \cite{edmonds:65a} and its
variants \cite{kolmogorov:09}. However, this approach is still not ideal as it
operates on the complete graph of the $F$ frustrated plaquettes with $F(F-1)$ edges,
and the edge weights need to be computed in a preparatory step using a suitable
approach such as Dijkstra's algorithm \cite{gibbons:book} before attacking the
matching problem. This approach allows one to study systems of a typical maximum linear
size of $L \approx 500$ \cite{hartmann:01a}.

As was shown more recently \cite{thomas:07,gregor:07}, an alternative mapping is
significantly more efficient as it operates on a sparse graph with similar
connectivity as the original lattice. It relies on an auxiliary graph that replaces
each vertex of the lattice by a complete graph $K_4$ of four nodes, also known as
Kasteleyn city. Edge weights are set to $J_{ij}$ for the original edges and to zero
for the internal $K_4$ bonds. The solution of a matching problem on this graph then
corresponds to a set of closed loops on the original lattice, separating domains of
opposite spin orientations, cf.\ the illustration in Fig.~\ref{fig:matching}. The
configuration of minimum weight corresponds to a ground state of the spin-glass
sample. Due to the sparsity of the auxiliary graph, significantly larger systems can
be studied with this method as compared to the one proposed in \cite{bieche:80a}, and
calculations for square lattices up to $L = 10\,000$ have been reported in
Ref.~\cite{khoshbakht:17a}.

\begin{figure}[tb!]
  \begin{center}
    \includegraphics[width=0.45\columnwidth]{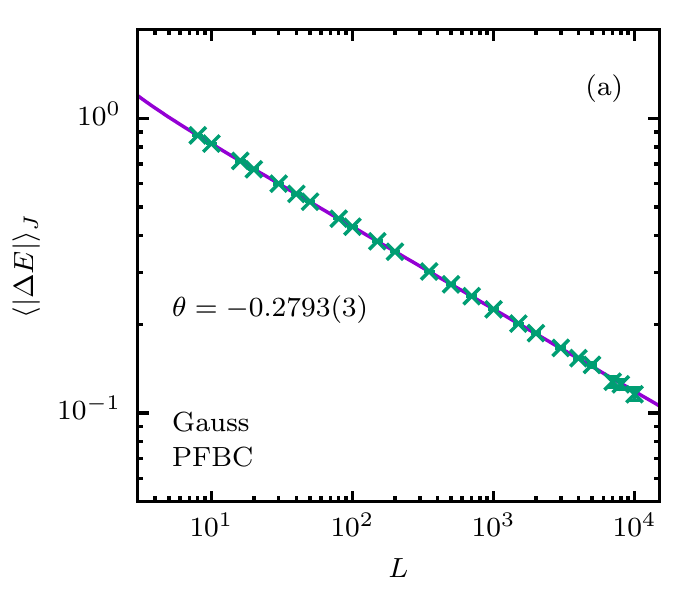} \hspace*{0.25cm}
    \includegraphics[width=0.45\columnwidth]{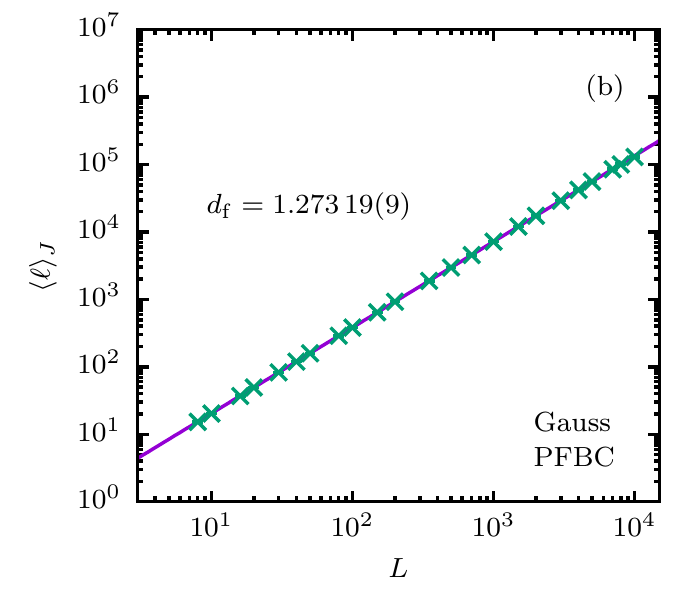}\\
    \caption{%
      (a) Scaling of defect energies $E_\mathrm{def}$ for the Gaussian EA model on
      the square lattice with periodic boundaries in $x$ direction and free
      boundaries in $y$ direction. The line shows a fit of the data to the functional form
      $E_\mathrm{def} = A_\theta L^\theta+C_\theta/L^2$
      \cite{khoshbakht:17a}. (b) Scaling of the domain-wall length between periodic
      and antiperiodic BCs. The line corresponds to a fit of the functional form
      \eqref{eq:fractal-dimension} to the data.}
    \label{fig:defect-scaling}
  \end{center}
\end {figure}

\begin{figure}[tb!]
  \begin{center}
    \raisebox{0.35cm}{\includegraphics[width=0.4\columnwidth]{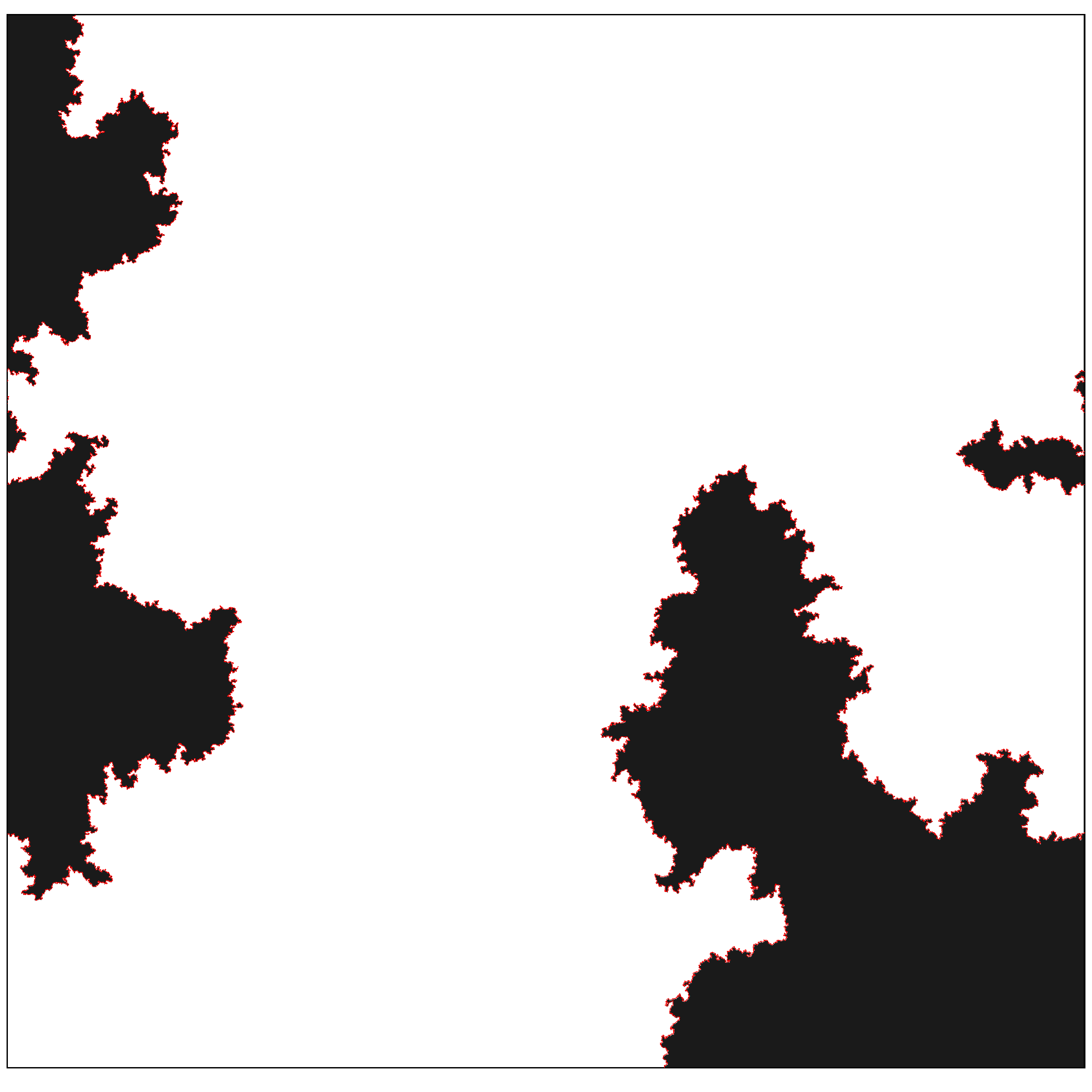}}  \hspace*{0.5cm}
    \includegraphics[width=0.4\columnwidth]{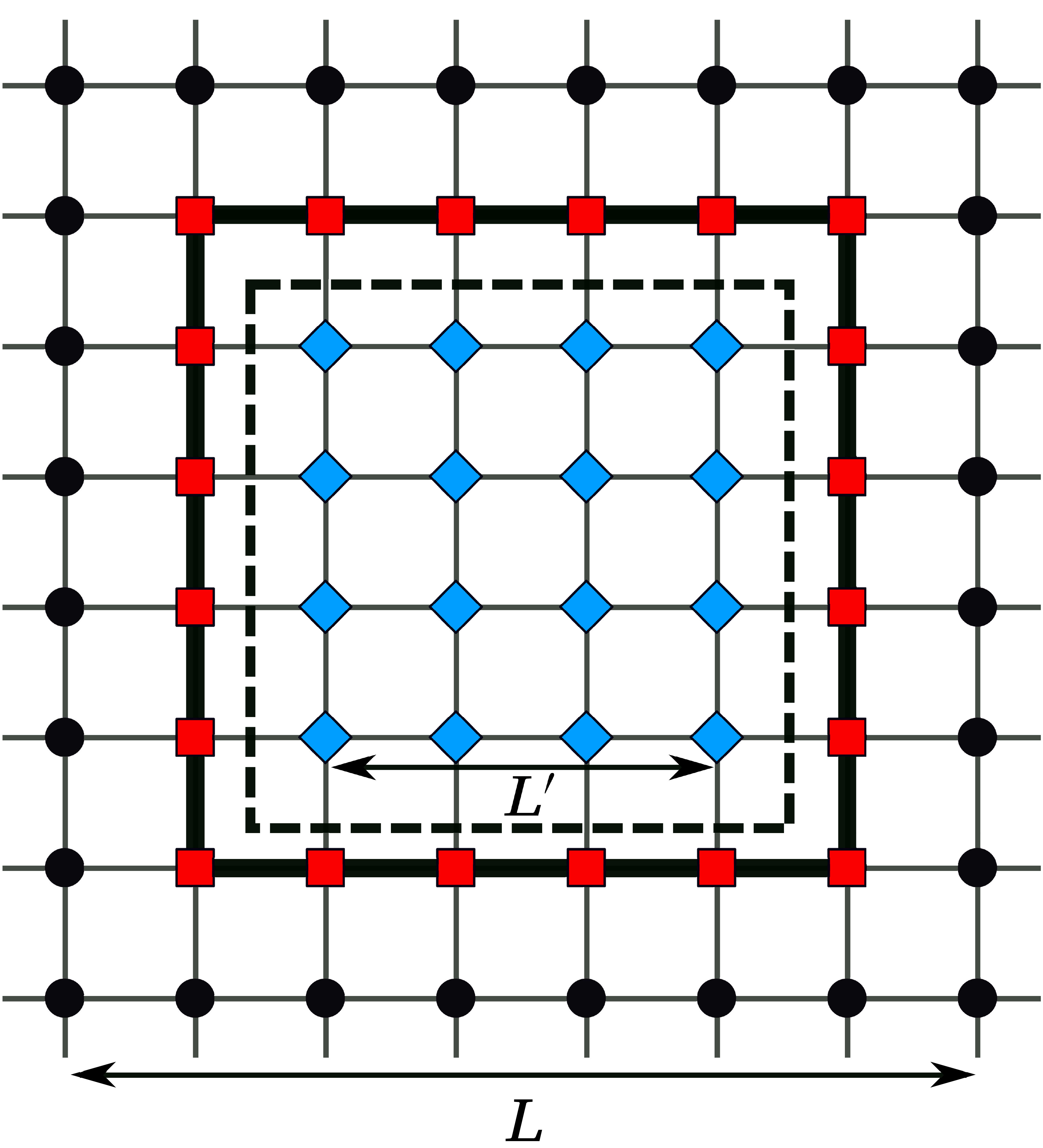}
  \end{center}
  \caption{Left: The domain wall separating regions of equal and opposite
    ground-state configurations for a specific disorder sample considered with
    periodic and with antiperiodic boundaries. Right: Setup used for the windowing
    technique to compute ground states for samples with fully periodic boundary
    conditions.}
  \label{fig:windowing}
\end{figure}

A widely applied method for studying the excitations out of the ground states that
take a central role in the theory of the spin-glass phase, consists of systematic
modifications of boundary conditions (BCs). It is argued that the defect energy
connected to a change from periodic to antiperiodic BCs in one direction,
\begin{equation}
  \label{eq:defect-energy}
  E_\mathrm{def} = |E_\mathrm{P}-E_\mathrm{AP}|,
\end{equation}
can act as a proxy for a typical low-energy excitation of the system \cite{banavar:82a}. Here,
$E_\mathrm{P}$ and $E_\mathrm{AP}$ refers to the ground-state energy for periodic and
antiperiodic BCs, respectively. One expects that $E_\mathrm{def}$ scales 
as \cite{bray1984,mcmillan1984},
\begin{equation}
  \label{eq:defect-scaling}
  E_\mathrm{def} \sim L^\theta
\end{equation}
with the {\em spin-stiffness exponent\/} $\theta$, where $\theta > 0$ should indicate
the stability of the spin-glass phase at non-zero temperatures, while for
$\theta < 0$ the transition temperature $T_\mathrm{SG} = 0$ and $\theta = -1/\nu$
governs the divergence of the spin-glass correlation length as $T\to T_\mathrm{SG}$
\cite{bray:87a}. For Gaussian exchange couplings $J_{ij}$ one finds a stiffness
exponent $\theta \approx - 0.3$ \cite{hartmann:01a}, with the most accurate estimate
being \cite{khoshbakht:17a},
\begin{equation}
  \label{eq:theta-value}
  \theta = -0.2793(3).
\end{equation}
This is illustrated in Fig.~\ref{fig:defect-scaling}(a) showing the scaling of defect
energies over a wide range of system sizes. The change of boundary conditions induces
a domain-wall defect that spans the system; a typical configuration of the overlap
between the ground states for periodic and antiperiodic BCs is shown in the
left panel of Fig.~\ref{fig:windowing}. The boundary of the flipped domain is a
fractal curve, and the domain-wall length is hence expected to show fractal scaling
of the form
\begin{equation}
  \label{eq:fractal-dimension}
  \langle \ell\rangle_J  = A_\ell L^{d_\mathrm{f}}.
\end{equation}
As is illustrated in Fig.~\ref{fig:defect-scaling}(b), this is indeed borne out in
the data to high accuracy, and the fractal dimension is estimated as $d_\mathrm{f} =
1.27319(9)$.

For technical reasons the matching approach can only handle samples on planar graphs,
i.e., lattices with periodic boundary conditions in at most one direction
\cite{bieche:80a}. More precisely, runs for systems with fully periodic boundaries
yield the same result for periodic and for antiperiodic BCs in each of the two
directions, such that the configuration returned is a ground state for one out of
four possible BCs. As pointed out in Ref.~\cite{thomas:07}, the approach hence
effectively optimizes over BCs as well as spin variables. To circumvent this problem
and allow treatment of systems with fully periodic BCs with the resulting smaller
scaling corrections, one may use a windowing technique as illustrated in the right
panel of Fig.~\ref{fig:windowing}: since MWPM can be used to find exact ground states
for planar graphs, in order to treat a periodic $L\times L$ system one applies it to
a square subset of edge length $L-2$ (``window'') while keeping the relative
orientation of the outside spins fixed. Randomly displacing the window location over
the (periodic) lattice, repeated applications of the window optimizations lead to a
quick convergence of the result. This prescription results in a stochastic algorithm,
whose success probability can be arbitrarily improved by using $m$ independent runs,
\begin{equation}
  P_s(\{J_{ij}\})=1-[1-P_{n}(\{J_{ij}\})]^m.
  \label{eq:success}
\end{equation}
Numerically, one finds that the number of required repetitions for a given success
probability is independent of system size, such that the computational complexity of
the algorithm remains the same as that of the MWPM approach for planar graphs, which
scales as $L^\kappa$ with $\kappa \approx 2.2$ \cite{khoshbakht:17a} for the Blossom
V algorithm \cite{kolmogorov:09}.

The system with bimodal couplings can also be studied with matching techniques. Here,
one finds no asymptotic decay of $E_\mathrm{def}$, but a convergence to a positive
limiting value as $L\to\infty$ \cite{hartmann:01a}. While this was initially taken as
evidence for a lower-critical dimension $d_l = 2$, it was later on realized that it
is rather a signature of an additional zero-temperature renormalization-group fixed
point that is not relevant for the physics at non-zero temperatures \cite{joerg:12}
and, instead, there is evidence for $d_l = 2.5$ \cite{boettcher:06,maiorano2018support}.
A signature of this system is the extensive degeneracy of the ground state, leading to a
finite ground-state entropy. This creates a difficulty for the matching approach as it
does not pick the individual ground states with equal probabilities. In its simplest form,
it is deterministic and will hence always return the same ground state. Some simple
randomizations through changing the order of considering bonds or adding some noise
onto the couplings improve on this, but still lead to biased sampling methods
\cite{khoshbakht:17a}. Unbiased sampling can be achieved via a suitably constructed
Monte Carlo sampling technique in the ground-state manifold, thus allowing for
estimates of the domain-wall fractal dimensions and related quantities
\cite{khoshbakht:17a}.

Matching techniques can also be used to sample other excitations than the domain-wall
perturbations induced by a change of boundary conditions. Different types of
droplet-shaped excitations can be studied by fixing the relative orientation of spins
through ``hard'' bonds \cite{hartmann:03a,hartmann:08}. Also, while matching is the
most widely used technique for this problem, an alternative approach based on a
calculation of the partition function through the use of Pfaffians allows to also study
finite-temperature properties exactly and in polynomial time
\cite{blackman:91,saul:93,galluccio:00}. Recently, such methods have been extended to
also enable the study of correlation functions and further related properties
\cite{thomas:09,thomas:11,thomas:13}.

\section{The assignment problem}\label{sec:assignment}
In this section we will discuss some new results in the {\em assignment problem}.
In order to define the question in its simplest version we have to consider a square matrix of size $n$, $W:=\{w_{ij}\}_{i,j=1}^n$, with real entries. For each permutation  $\pi$ in the symmetric group ${\mathcal S}_n$ consider the matrix $\Pi:=\{\pi_{ij}\}_{i,j=1}^n$ with entries zero or one, such that 
\begin{equation}
\pi_{ij} = \delta_{j,\pi(i)} =
\begin{cases}
1 & \hbox{\rm if\, } \pi(i) = j \\
0 & \hbox{\rm elsewhere}
\end{cases}
\end{equation}
and ask for a permutation of the columns of $W$ in order to get a minimum of the trace, that is of the Hamiltonian 
\begin{equation}
{\mathcal H}(\Pi,W) := \trace \left(W^T\cdot \Pi \right) =  \sum_{i,j=1}^n w_{i j} \pi_{i j}  = \sum_{i,j=1}^n w_{i j} \delta_{j,\pi(i)} = \sum_{i=1}^n w_{i, \pi(i)}   \, .
\end{equation}
Without loss of generality we can take the entries of $W$ to be nonnegative: $w_{ij}\ge 0$. Indeed, in {\em combinatorial optimization} they are usually referred to as {\em costs}. In different words, this is the classical {\em matching problem} on the {\em bipartite complete graph} ${\mathcal K}_{n,n}$.
The solution of the problem is  a permutation $\pi^*$, with corresponding matrix $\Pi^*(W)$ , that realizes the minimum cost and brings to the determination of
\begin{equation}
{\mathcal H}^*(W) := \min_{\pi \in {\mathcal S}_n} {\mathcal H}(\Pi,W) = {\mathcal H}\left(\Pi^*(W),W\right)
\end{equation}
the optimal value of the total cost. The optimal solution can be seen as the ground state of the Hamiltonian of a {\em disordered} system defined by the cost-matrix $W$, an analogy which can be made useful, for example, by {\em simulated annealing}~\cite{kirkpatrick1983}.

The introduction of a probability on the space of the possible costs allows the investigation of the properties of the {\em typical} solutions and the artillery from statistical physics shows its whole power~\cite{Mezard1987spin,Hartmann2006book,Mezard2009information}. In their seminal works M\'ezard and Parisi ~\cite{Mezard1985,Mezard1986,Mezard1986a} (but see also~\cite{Orland1985}) could solve, by using the replica trick,  the matching, the assignment and the Travelling Salesman Problem in the case in which the entries $w_{ij}$ are independent random variables, equally distributed, with a probability distribution density of the form
\begin{equation}
\rho(w) = w^r \sum_{k=0}^\infty \eta_k w^k
\end{equation}  
with $\eta_0\neq 0$.
More precisely, in the asymptotic limit of an infinitely large size $n$, replica symmetry is not broken, and the optimal solution is determined by the application of the saddle-point method as the solution of an integral equation in which only the parameter $r$ enters. In particular
\begin{equation}
E_n := \mathbb{E} \left[{\mathcal H}^*(W)\right] \sim \,n^{1-\frac{1}{r+1}}\,,
\end{equation}
where the expectation value is taken on the possible weights and with $\sim$ we indicate that both sides of the relation scale with large $n$ in the same way so that their ratio converges in the limit of an infinite number of points. 
For a recent work in which the first finite-size corrections are reconsidered after~\cite{Mezard1987,Parisi2002} and the extension to non-integer value of the parameter $r$ see~\cite{Caracciolo:168}.

A new class of problems arises when the vertices of the graph ${\mathcal K}_{n,n}$ are identified with points in $\Omega\subset {\mathbb R}^d$ seen as a subset of a Euclidean space. We thus have two sets of points of cardinality $n$: we denote them as the red points $\mathcal R$, respectively blue points $\mathcal B$, with positions $x_i\in {\mathbb R}^d$, respectively $y_i\in {\mathbb R}^d$, with $i\in\{1,\dots, n\} = [n]$. Then the cost of the assignment of the $i$-th red point with the $j$-th blue point is assumed to be a function of the Euclidean distance between the two points $|x_i-y_j|$. We shall restrict to the cases
\begin{equation}
w_{ij} = f(|x_i-y_j|) = |x_i-y_j|^p	
\end{equation}
parametrized by the real number $p\in {\mathcal R}$. Let us introduce the {\em empirical} probability measures associated to the two sets of points
\begin{equation}
\rho_{\mathcal R}(x) = \frac{1}{n} \sum_{i=1}^n \delta(x-x_i) \, ; \qquad \rho_{\mathcal B}(y) = \frac{1}{n} \sum_{i=1}^n \delta(y-y_j)
\end{equation}
and look, for $p\ge 1$,  at the  $p$-th Wasserstein (elsewhere associated to the names of  Kantorovich and Rubinstein) distance of two probability measures, that  is at the variational problem 
\begin{equation}
W_p(\rho_1, \rho_2) := \left(\inf_{\gamma\in\Gamma(\rho_1, \rho_2) }\int  dx \, d y\, \gamma(x,y) \, |x-y|^p \right)^\frac{1}{p}\,,
\end{equation}
where $\Gamma(\rho_1, \rho_2)$ is the set of measures on $\Omega\times\Omega$ with marginals $\rho_1$ and $\rho_2$. This is nothing but the {\em optimal transport} (or Monge-Kantorovich) problem of a unit mass distributed according to $\rho_1$ to a distribution $\rho_2$~\cite{Villani2008,Ambrosio2003a}. A function $\gamma(x,y)$ defines a {\em transportation plan}, indeed, the marginality conditions
\begin{equation}
\int d x\, \gamma(x,y) = \rho_2(y) \, ;\qquad \int d y\, \gamma(x,y) = \rho_1(x)
\end{equation}
constraining the mass moved into the point $y$ and from the point $x$ are what is required.
In the case of the two empirical probability measures the possible transportation plans reduce to the set of permutations and therefore their Wasserstein distance is simply related to the optimal cost~\cite{Brezis}
\begin{equation}
{\mathcal H}^*(W) = n \, W_p^p(\rho_{\mathcal R}, \rho_{\mathcal B}).
\end{equation}
A random matrix, whose elements depend on the Euclidean distance between points randomly distributed in space is called {\em Euclidean}. The spectra of Euclidean random matrices have been studied in~\cite{MPZ}.

In order to fix the ideas, let us consider the case in which $\Omega = [0,1]^d$, periodic boundary conditions are chosen, so that we are really on a torus of unit volume and the positions of red and blue points are taken at random with flat probability. A natural length-scale is therefore $n^{-\frac{1}{d}}$ and one can simply assume that for each red point there is at least a blue point to match in a ball of radius of order $n^{-\frac{1}{d}}$, so that we can guess that 
\begin{equation}
E_n \sim n^{1-\frac{p}{d}}\, .
\end{equation}
But it is well known~\cite{Ajtai} that this can be true only in $d \ge 2$. It was proven that in $d=2$
\begin{equation}
E_n  \sim n \, \left(\frac{\log n}{n} \right)^{\frac{p}{2}}
\end{equation}
a logarithmic violation appears. A much more detailed analysis is possible in $d=1$~\cite{Caracciolo:159} where an intriguing relation with Brownian processes is explored~\cite{Caracciolo:160}. It is shown that, for a generic distribution probability $\rho$ with cumulative $\Phi$, the average total cost is
\begin{equation}
E_n = n^{1-\frac{p}{2}} \frac{2^p}{\sqrt{\pi}} \Gamma\left(\frac{p+1}{2}\right) \int_0^1 d x \frac{[\Phi(x) (1 - \Phi(x))]^\frac{p}{2}}{\rho^{p-1}(x)} + o(n^{-\frac{p}{2}})
\end{equation}
at least when the integral is convergent. Otherwise an anomalous scaling emerges~\cite{Caracciolo:172}. In the case of open boundary conditions, with flat distribution, the average cost can be evaluated even for finite size by means of Selberg integrals~\cite{Caracciolo:177}
\begin{equation}
E_n = n\, \frac{ \Gamma\left(1+\frac{p}{2}\right) }{p+1} \frac{\Gamma(n+1)}{\Gamma\left(n+1+\frac{p}{2}\right)}\, .
\end{equation}
Also the case $p\leq 0$ has been studied~\cite{Caracciolo:169}, where, instead, for flat distribution
\begin{equation}
\lim_{n\to\infty}\frac{E_n}{n} = \frac{1}{2^p}\,.
\end{equation}
When $0<p<1$ the cost function becomes concave. In~\cite{Caracciolo:180} it is argued that
\begin{equation}
E_n \sim
\begin{cases}
n^{1-p} & \hbox{\rm for\,} 0<p<\frac{1}{2}\,,\\
\sqrt{n} \log n & \hbox{\rm for\,} p=\frac{1}{2}\,,\\
\sqrt{n} & \hbox{\rm for\,} \frac{1}{2}<p<1\,, \\
\end{cases}
\end{equation}
so that a change in the phase diagram occurs at $p=\frac{1}{2}$.

An amusing exact result has been obtained for the flat distribution in $d=2$ for the particular value $p=2$, in~\cite{Caracciolo:158, Caracciolo:162}, that is 
\begin{equation}
\lim_{n\to\infty}\frac{E_n}{\log n} = \frac{1}{2 \pi}.
\end{equation}
This has been rigorously proven in~\cite{Ambrosio2016} (see also~\cite{Ambrosio2019}  for improvements). Let us follow the field theoretic approach introduced in~\cite{Caracciolo:163} and let us introduce the vector {\em transport field} $\mu(x)$ which joins the red point in $x$ to the blue point in $y=x+\mu(x)$ and consider the Lagrangian
\begin{equation}
{\mathcal L}[\mu,\phi] := \frac {1}{2} \int \mu^2(x) \rho_{\mathcal R}(d x) + \int \left[ \phi(x+\mu(x)) \rho_{\mathcal R}(d x) - \phi(x) \rho_{\mathcal B}(d x)\right]
\end{equation}
to be minimized, where the scalar field $\phi(x)$ is a Lagrangian multiplier which implements a matching between blue and red points. In the limit of a large number of points $n$, when the red and blue points are extracted with the same distribution $\rho$ so that $\delta \rho := \rho_{\mathcal R} - \rho_{\mathcal B}$ and the optimal $\mu$ goes to zero and we expect a good approximation by using the only the quadratic terms in the Lagrangian, that is
\begin{equation}
{\mathcal L}[\mu,\phi] :=  \int \left[\frac {1}{2} \mu^2(x) + \mu(x)\cdot \nabla \phi(x) \rho(d x) \right] +  \int  \phi(x) \delta \rho(d x)\,,
\end{equation}
which has Euler-Lagrangian equations
\begin{equation}
\mu = -\nabla \phi \, , \qquad \nabla \cdot [\rho\, \mu] = \delta\rho\,,
\end{equation}
revealing a strict analogy with an electrostatic problem where $\mu$ plays the role of the electric field, $\phi$ is the scalar potential and indeed is the Lagrangian multiplier which implements the Gauss law, red and blue points have opposite unit charge, being null the total charge, while $\rho$ is the dielectric function of a linear dielectric medium. As a consequence 
\begin{equation}
- \nabla \cdot [\rho\, \nabla \phi] = \rho
\end{equation}
is solved by means of the classical Green's function $G_\rho(x,y)$ of the operator $-\nabla \cdot [\rho\, \nabla \bullet]$, so that an explicit approximate solution at fixed disorder is given by
\begin{equation}
\mu(x) = \int \nabla_x G_\rho(x,y)\, \delta\rho(d y) \, .
\end{equation}
After averaging over disorder, in the simple case of a flat measure (see~\cite{benedetto2019} for non-constant case) we get
\begin{equation}
\lim_{n\to\infty} E_n(\Omega) = -2 \trace \Delta^{-1}_\Omega\,,
\end{equation}
where the Laplacian $\Delta_\Omega$ is defined on the domain $\Omega$. This formula is correct in $d=1$ but it simply provides a divergence on both sides for $d>1$. It is an ultraviolet divergence which has been introduced by the linearization in the infinite number of modes, but the approximation cannot be true at very short distances. In the exact non-linear theory higher modes are cut off and we expect that
\begin{equation}
E_n(\Omega) = 2 \int_{0^+}^\infty \frac{F\left(\frac{\lambda}{n}\right)}{\lambda} \,  d \mathcal{N}_\Omega(\lambda)\,,
\end{equation}
 where  $\mathcal{N}_\Omega(\lambda)$ is number of the eigenvalue less than $\lambda$ for the Laplace operator and the unknown function $F$ interpolates between 1 for $\lambda \lesssim n$ and 0 for $\lambda \gtrsim n$.

By the Weyl law~\cite{Ivrii2016} on the asymptotics of the eigenvalue counting function for the Laplace-Beltrami operator we know that, for a 2-$d$ manifold with unit volume (under Neumann boundary conditions which are appropriate for our problem)
\begin{equation}
\mathcal{N}_\Omega(\lambda) = \frac{1}{4 \pi} \left( \lambda + \sqrt{\lambda}\,  |\partial \Omega| \right) + o\left(\sqrt{\lambda}\right)\,.
\end{equation}
In $d=2$ it is easy now to evaluate the leading logarithmic singularity in the number of points and in agreement with~\cite{ Ambrosio2018} we expect that
\begin{equation}
E_n(\Omega) = \frac{1}{2 \pi} \log n + 2 c_*(n) + 2 c_\Omega + o(1)\,,
\end{equation}
where $c_*(n) = O(\log n)$ is a universal function not depending on $\Omega$. As a further consequence,
\begin{equation}
\lim_{n\to \infty} \left[E_n(\Omega)-E_n(\Omega^\prime)\right]  = 2 \lim_{n\to \infty} \int_{0^+}^\infty \frac{F\left(\frac{\lambda}{n}\right)}{\lambda} \, \left[ d \mathcal{N}_\Omega(\lambda) -  d \mathcal{N}_{\Omega^\prime}
 (\lambda)\right]\,,
\end{equation}
but the r.h.s. is convergent even in the absence of regularization thus 
\begin{equation}
\lim_{n\to \infty} \left[E_n(\Omega)-E_n(\Omega^\prime)\right]  = 2  \int_{0^+}^\infty \frac{d \mathcal{N}_\Omega(\lambda) -  d \mathcal{N}_{\Omega^\prime}(\lambda)}{\lambda}\,,
\end{equation}
an expression that has been tested in~\cite{Caracciolo:182} on various 2$d$ manifolds, by using both the Green's function method as other classical tools as the Dedekind's limit formulas, thus confirming the predictions of the field theoretic approach.

For a similar method applied to a more general context see~\cite{Koehl2019,Koehl2021,Koehl2021b}.

Once more this relatively simple combinatorial optimization problem is revealing intriguing and fruitful connections with so many different research fields.

\newpage
\bibliographystyle{ws-rv-van}
\bibliography{biblio1}

\end{document}